# Dynamic Sublimation Pressure and the Catastrophic Breakup of Comet ISON

Jordan K. Steckloff[a], Brandon C. Johnson[b], Timothy Bowling[c], H. Jay Melosh[a,c,d], David Minton[c], Carey M. Lisse[e], and Karl Battams[f]

[a]Purdue University, Department of Physics and Astronomy, 525 Northwestern Avenue, West Lafayette, IN 47907

[b]Massachusetts Institute of Technology, Department of Earth, Atmospheric, and Planetary Sciences, 77 Massachusetts Avenue, Cambridge, MA, 02139

[c]Purdue University, Department of Earth, Atmospheric, and Planetary Sciences, 550 Stadium Mall Drive, West Lafayette, IN 47907

[d]Purdue University, Department of Aeronautical and Astronautical Engineering, 701 W. Stadium Avenue, West Lafayette, IN 47907

[e]John Hopkins University – Applied Physics Laboratory, 11100 Johns Hopkins Road, Laurel, MD, 20723

[f]Naval Research Laboratory, 4555 Overlook Avenue, Washington, D.C., 20375

Corresponding Author: Jordan Steckloff

## Abstract

Previously proposed mechanisms have difficulty explaining the disruption of Comet C/2012 S1 (ISON) as it approached the Sun.  We describe a novel cometary disruption mechanism whereby comet nuclei fragment and disperse through dynamic sublimation pressure, which induces differential stresses within the interior of the nucleus.  When these differential stresses exceed its material strength, the nucleus breaks into fragments. We model the sublimation process thermodynamically and propose that it is responsible for the disruption of Comet ISON.  We estimate the bulk unconfined crushing strength of Comet ISON's nucleus and the resulting fragments to be 0.5 Pa and 1–9 Pa, respectively, assuming typical Jupiter Family Comet (JFC) albedos.  However, if Comet ISON has an albedo similar to Pluto, this strength estimate drops to 0.2 Pa for the intact nucleus and

0.6-4 Pa for its fragments. Regardless of assumed albedo, these are similar to previous strength estimates of JFCs. This suggests that, if Comet ISON is representative of dynamically new comets, then low bulk strength is a primordial property of some comet nuclei, and not due to thermal processing during migration into the Jupiter Family.

# 1 Introduction

On November 12, 2013 sungrazing comet C/2012 S1 (ISON) unexpectedly disrupted into fragments. This occurred at a heliocentric distance of 145 solar radii ($R_\odot$) (0.68 AU), prior to reaching perihelion (Combi et al. 2014; Boehnhardt et al. 2013; Steckloff et al. 2015). Subsequent disruption events occurred on November 21 and 26 at 88 $R_\odot$ (0.41 AU) and 36 $R_\odot$ (0.17 AU) respectively (Knight & Battams, 2014; Steckloff et al. 2015). While there is nothing seemingly special about these heliocentric distances, currently known sungrazing comet disruption mechanisms seem inadequate to explain ISON's demise. ISON's disruptions occurred much too far from the Sun to have been caused by ablation or chromospheric impact, which disrupt nuclei within a heliocentric distance ($q$) of 1.01 $R_\odot$ (Brown et al. 2011). Tidal stresses can disrupt the nucleus only within the fluid Roche Limit ($q < \sim 2\ R_\odot$) (Knight & Walsh, 2013). Additionally, ISON's effective radius of ~600-700 m (Delamere et al. 2013; Lamy et al. 2014) was too large to have lost all its ice through complete sublimation and then disintegrated, a process that may only disrupt nuclei less than ~200-350 m in radius (Knight & Walsh, 2013; Sekanina, 2003). Finally, ISON's 10.4 hour rotation period at 210 $R_\odot$ on November 1 (Lamy et al. 2014) was too long for nongravitational torques to spin the body up to fragmentation (~2.2 hour period) (Pravec et al. 2006) by the time it reached 145 $R_\odot$ less than 2 weeks later on November 13 (Samarasinha & Mueller, 2013). However, it has been implied that sublimating gases are linked to the disruption of sungrazing comets (Sekanina 2003). Here we introduce a new break-up mechanism that readily explains Comet ISON's series of disruptions.

As illustrated in *Figure 1,* gas sublimating on the sunward side of the nucleus transfers momentum to the nucleus, exerting a dynamic sublimation pressure on its illuminated hemisphere. The sublimation pressure on the surface generates differential stresses within the nucleus that may exceed ISON's material strength, ultimately disrupting the comet into fragments (Brown et al. 2011; Borovička et al. 2013). Based on the timing of disruption events we can estimate the bulk unconfined crushing strength of Comet ISON's nucleus.

## 2  Theory/calculation

Investigating our proposed disruption mechanism requires an accurate computation of the sublimation pressure (itself a function of both thermal gas velocity and mass loss rate) acting at the surface of the nucleus as a function of heliocentric distance. Previous computations of cometary sublimation rely heavily upon either empirical fits to observed volatile mass loss rates (e.g. Marsden et al. 1973; Cowan & A'Hearn 1979; Sekanina, 1992), or on the theoretical dependence of mass loss rates on temperature (Delsemme & Swings, 1952) rather than the dependence of sublimation pressure on heliocentric distance. We choose instead to construct a versatile thermodynamic model of the sublimation pressure acting upon a cometary surface. In our calculations, the heliocentric dependence of the sublimation pressure of a particular volatile species is fully described by six known quantities: heliocentric distance ($r_{helio}$), molar mass ($m_{molar}$), heat of sublimation ($L$), sublimation coefficient ($\alpha$) and a laboratory measurement of vapor pressure ($P_{ref}$) at a known temperature ($T_{ref}$).

Comets consist of intimate mixtures of refractory materials (silicates, metal sulfide dust, organics) and volatile ices (primarily $H_2O$, $CO_2$, and CO [Bockleé-Morvan et al. 2004]). The phase-change behavior of mixtures of volatiles can be significantly more complicated than that of a single, pure volatile species. In particular, if cometary CO is mostly trapped within amorphous $H_2O$ ice, then the release of significant quantities of CO may require the amorphous $H_2O$ ice to crystallize (Bar-Nun et al. 2013), which is a highly exothermic and potentially explosive phase transition (Mastrapa et al., 2013). Moreover, the presence of amorphous ice in comets is contentious (Lisse et al. 2013). However, Comet ISON's CO content is only a few percent of its $H_2O$ content (Weaver et al., 2014) and produced an order of magnitude less $CO_2$ than $H_2O$ (McKay et al. 2014). Therefore, we may assume that the sublimation pressure acting on Comet ISON's surface is dominated by the sublimation of pure $H_2O$ ice, which avoids the complications of the sublimation of mixed materials and species more volatile than $H_2O$ ice. However, we include the cases in which pure $CO_2$ and CO ice sublimates for the sake of comparison, which admittedly ignores the complications of how one would trap significant quantities of CO ice in the first place.

Typical bond albedos measured for Jupiter Family Comet (JFC) nuclei are very low (0.03-0.06) (Li et al. 2013a; Li et al. 2013b; Capaccioni et al. 2015), and when JFCs approach the Sun, most of the incident radiation (94-97%) is absorbed at the surface and drives the sublimation of volatile ices (an active comet's dominant cooling mechanism).

We explore the case in which Comet ISON's albedo is similar to that of JFCs, and assume that all incident radiation is absorbed (bond albedo of 0). However, because dynamically new comets have never been thermally processed by the Sun, it is plausible that their surfaces are significantly richer in ices than JFCs, which could lead to a much higher albedo. Moreover, there are no high-resolution observations of dynamically new comet nuclei, which would constrain their albedos. We therefore also explore the case in which Comet ISON has a bond albedo of 0.5, which is similar to that of the dwarf planet Pluto.

Observations of JFC nuclei suggest that cometary thermal inertia is very low (Gulkis et al. 2015; Davidsson et al. 2013; Groussin et al. 2013; Lisse et al. 2005; Lamy et al. 2008), meaning that little daytime heat is stored by the surface to be later released when it rotates into night. This naturally explains their highly asymmetric dayside-nightside distribution of sublimating gases (Feaga et al. 2007; Gulkis et al. 2015). Similarly, Comet ISON's activity is concentrated on its illuminated hemisphere (Li et al. 2013c). Since cometary activity is driven by volatile sublimation, we assume that effectively all volatile emission occurs on Comet ISON's illuminated hemisphere, causing a sublimation pressure that only acts on the illuminated parts of its nucleus. Indeed, it has been known for decades that nongravitational forces push predominantly on the sunward hemispheres of comet nuclei (Marsden et al. 1973). While observations show that the unilluminated side of comet nuclei can emit volatiles, emission on the unilluminated side is usually less than half of the emission of the illuminated side (Feaga et al. 2007; Gulkis et al. 2015). Therefore, our sunward emission assumption is valid for our purpose of obtaining an order of magnitude estimate of ISON's strength.

While the nuclei of highly thermally evolved comets (like JFCs) emit dust and gas from only a small fraction of their surfaces (Ververka et al. 2013, Samarasinha & Mueller, 2013), ISON's high $H_2O$ production rate prior to disruption suggests that nearly the entire surface of its nucleus was active (Combi et al. 2014), consistent with a thermally primitive, dynamically new comet. This implies that volatile ices are located within the thermal skin depth of the comet's surface. We therefore assume that volatile ices sublimate from the entire illuminated surface of ISON, and that a negligible amount of incident solar energy is thermally radiated into space from a mantle of material covering the volatile ices.

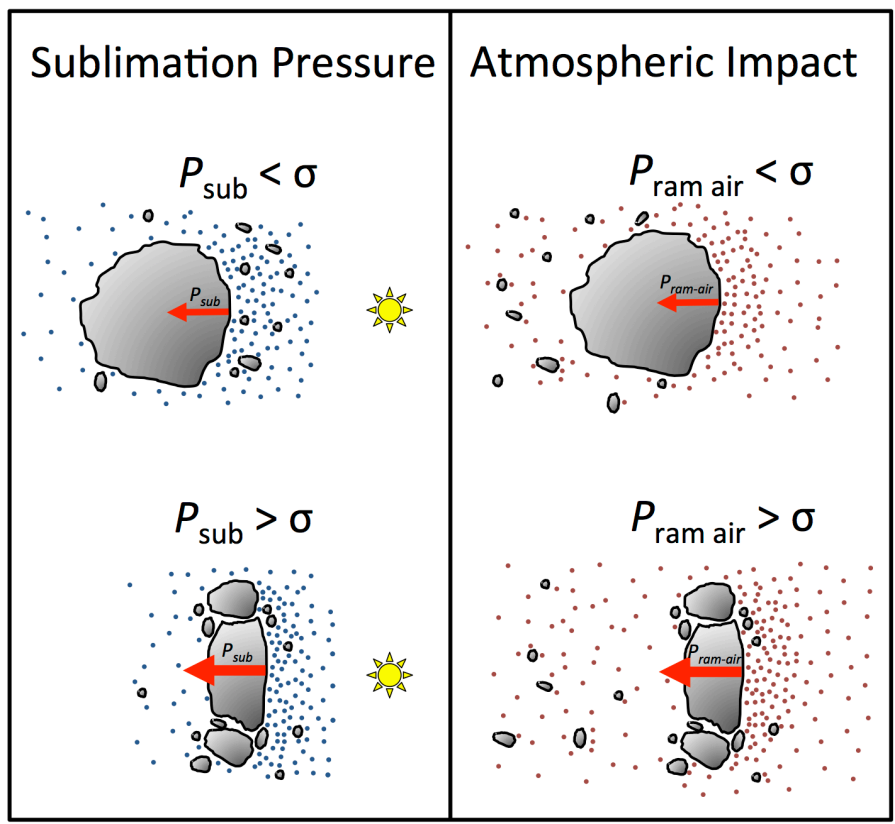

**Figure 1:** *Schematic of Dynamic Sublimation Pressure Disruption Mechanism and Comparison to Atmospheric Impact.* (**Left**) We assume that the dynamic pressure is zero on the dark side of the nucleus, while the peak dynamic pressure on the illuminated side

($P_{sub}$) becomes comparable to the unconfined static crushing strength of the nucleus (σ). When $P_{sub}$ exceeds σ, the nucleus disrupts catastrophically. (**Right**) This is analogous to the nucleus impacting a planetary atmosphere. A ram pressure ($P_{ram}$) builds up on the leading edge of the nucleus as it travels through the atmosphere. If $P_{ram}$ exceeds σ, then the nucleus breaks up into fragments (Borovička et al. 2013).

The dynamic pressure exerted by sublimating volatiles on the surface of the nucleus is equal to the momentum flux of the departing material, and is computed by multiplying the volatile's mass flux by its thermal velocity. Assuming that volatile ices are at or near the surface, we estimate Comet ISON's volatile mass flux by equating the absorbed solar energy to the energy required to sublime each ice species, as first described by Fred Whipple (Whipple, 1950). We assume that volatile ices and refractory materials are intimately mixed, such that heat is rapidly transferred from refractory materials to volatile ices. We ignore the amount of energy required to warm the ices from their initial low temperatures (perhaps 10 K for dynamically new comets such as ISON) to the equilibrium sublimation temperature. Such heating consumes less than ~10%, ~25%, and ~25% of the total incident solar energy for $H_2O$, $CO_2$, and CO ice respectively, and is therefore negligible for our order of magnitude estimates. For simplicity, we treat each volatile species individually, while acknowledging that multiple species may sublime simultaneously from different depths below the surface.

**2.1 Computing Mass Flux, Force, Temperature, and Sublimation Pressure**

The incident solar radiation intensity at the location of the comet is given by

$$I_{solar} = \frac{L_{solar}}{4\pi r_h^2} \qquad (1)$$

where $L_{solar}$ is the solar luminosity ( 3.846 x $10^{26}$ W), and $r_h$ is the heliocentric distance. We assume that all solar radiation incident upon an area element of the surface of the nucleus ($dA$) is used to overcome the latent heat of sublimation of these volatile ices (Whipple, 1950) to determine each species' mass flux

$$\dot{m} = (1-A)\frac{I_{solar}}{\lambda(T)}\cos\phi = (1-A)\frac{L_{solar}}{4\pi r_h^2 \lambda(T)}\cos\phi \qquad (2)$$

where $A$ is the albedo of the sublimating surface, $\lambda(T)$ is the temperature-dependent latent heat of sublimation of a volatile ice species and $\phi$ is the angle between the comet-Sun line and the vector normal to the area element (local phase angle). For a sphere, $\phi$ is equivalently the azimuth angle of the area element from the subsolar point. While the latent heat of sublimation for water is temperature-dependent, it varies so little over the temperature range of interest (Feistel & Wagner, 2007) that treating it as a constant makes a negligible difference in our results. We therefore assume that the latent heat of sublimation is a constant.

We determine the thermal velocity of the dominant sublimating volatile using the kinetic theory of gases. We assume that the speeds of sublimating gas molecules obey a Maxwell-Boltzmann distribution, where the mean of the magnitude of the molecule velocities escaping from a given area element ($dA$) is

$$v_{thermal} = \sqrt{\frac{8RT}{\pi m_{mol}}} \qquad (3)$$

where $m_{mol}$ is the molar mass of the species, $T$ is the gas temperature, and $R$ is the ideal gas constant. The gas diffusing through the cometary pores has a Knudsen number of Kn~$10^2$-$10^5$, which allows us to assume that the sublimating volatile molecules are sufficiently rarefied to be emitted from a porous regolith according to Lambert's cosine law (Gombosi, 1994, pp. 227-230). Thus, the number of molecules emitted in a particular direction from an area element ($d\dot{N}(\theta)$) is proportional to the cosine of the angle of that direction with respect to the vector normal to that area element

$$d\dot{N}(\theta) = \frac{\dot{N}_{dA}}{\pi} \cos\theta \tag{4}$$

where $\dot{N}_{dA}$ is the number flux of molecules through area element $dA$, and $\theta$ is the angle made with the vector normal to area element $dA$. We compute the net force on a given area element from sublimating gas molecules by multiplying this particle density distribution by both $v_{thermal}$ and the mass of a particle, and then integrate over all solid angles. Since the particle density distribution depends solely on the angle with respect to the vector normal to the area element, this computation is axisymmetric. Thus, the components of the force tangential to the surface of area element $dA$ cancel out, allowing us to consider only the component of the force normal to the surface. Integrating over all solid angles above the ground

$$F_{element} = \frac{2}{3} v_{thermal}\, \dot{m}\, dA \tag{5}$$

and the mass flux from the area element ($\dot{m}$) is

$$\dot{m} = \frac{m_{molar}}{N_{av}} \frac{\dot{N}_{dA}}{dA} \tag{6}$$

where $m_{molar}$ is the molar mass of the sublimating gas and $N_{av}$ is Avogadro's constant. Combining equations (2), (3), and (5)

$$F_{element} = \frac{2}{3}(1-A)\frac{L_{solar}}{4\pi r_h^2 \lambda} \sqrt{\frac{8RT}{\pi m_{mol}}} \cos\phi\, dA \tag{7}$$

We compute the appropriate temperature ($T$) in Equation (7) by joining the Langmuir-Knudsen (Langmuir, 1913) equation of sublimation rates with the Clausius-Clapyron relation of equilibrium partial pressure and temperature of an ideal gas

$$\dot{m} = \alpha(T)\sqrt{\tfrac{m_{mol}}{2\pi RT}}P(T) \tag{8}$$

$$\frac{dP}{dT} = \frac{P}{T^2}\frac{\lambda}{R} \tag{9}$$

where $\alpha(T)$ is the temperature-dependent sublimation coefficient (e.g. Gundlach et al. 2011) and $P(T)$ is the temperature-dependent partial pressure of the molecular species, which results in the following expression for the temperature as a function of the mass flux:

$$\dot{m} = \alpha(T)\sqrt{\tfrac{m_{mol}}{2\pi RT}}P_{ref}\, e^{\tfrac{\lambda}{R}\left(\tfrac{1}{T_{ref}}-\tfrac{1}{T}\right)} \tag{10}$$

where $P_{ref}$ and $T_{ref}$ are an experimentally measured reference pressure and temperature of the species. We use the empirical fit to the temperature dependence of the sublimation coefficient $\alpha(T)$ for $H_2O$ from Gundlach et al. (2011), which produces a small improvement in the computation of water's sublimation pressure over setting the sublimation coefficient to 1. We set the sublimation coefficient $\alpha(T)$ for all other species to 1. Combining equations (2) and (10)

$$(1-A)\frac{L_{solar}}{4\pi r_h^2 \lambda}\cos\phi = \alpha(T)\sqrt{\tfrac{m_{mol}}{2\pi RT}}P_{ref}\, e^{\tfrac{\lambda}{R}\left(\tfrac{1}{T_{ref}}-\tfrac{1}{T}\right)} \tag{11}$$

Note that this is a transcendental equation, which does not have an analytical solution. Thus, we solve for this temperature numerically. Lastly, since pressure is a force applied over an area, we rearrange equation (7) to describe the dynamic sublimation pressure exerted on the surface of a nucleus

$$P_{sub}(r_h,\phi) = \tfrac{2}{3}(1-A)\frac{L_{solar}}{4\pi r_h^2 \lambda}\sqrt{\tfrac{8RT}{\pi m_{mol}}}\cos\phi \tag{12}$$

We approximate a comet as a sphere, and plot the dependence of the dynamic sublimation pressure on the azimuth from the subsolar point ($\phi$) for the sublimation of $H_2O$ at heliocentric distances of 36 $R_\odot$, 88 $R_\odot$, and 145 $R_\odot$ (see *Figure 2*).

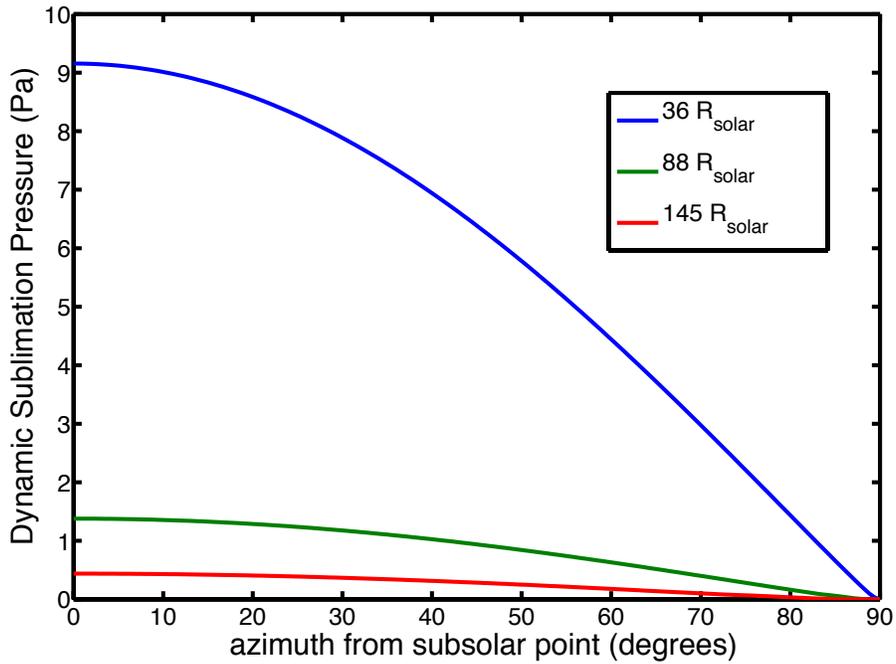

**Figure 2:** *Azimuthal Dependence of Dynamic Sublimation Pressure.* A plot of the azimuthal dependence of the dynamic sublimation pressure for three separate heliocentric distances for the case of a bond albedo of 0. Azimuthal angle is the angle between the subsolar point and the vector normal to the surface of an idealized, spherical nucleus. While the real nucleus is not necessarily spherical, it will have a subsolar point and a limb, where the dynamic sublimation pressures will be at a maximum and zero respectively. The differential stress that results from this pressure difference is ultimately responsible for fragmenting the nucleus.

## 2.2 Differential Stress

Computing differential stresses with ISON's nucleus is essential to our analysis, because differential stresses can lead to its disruption. A compressive differential stress will cause a brittle material to deform, but the material will remain intact deforming elastically as long as the differential stress remains below the material's strength. However, when the differential stress exceeds a brittle material's strength, the material will fail and fracture. In the case of a comet, when the dynamic sublimation pressure causes material failure, the nucleus will subsequently fragment.

Because the sublimation pressure drops to zero at a 90-degree azimuth from the subsolar point (the limb of the nucleus) and remains near zero on the unilluminated side, the maximum differential stress within the nucleus is similar in magnitude to the sublimation pressure at the subsolar point (the maximum sublimation pressure). Therefore, when we compute the dynamic sublimation pressure at the subsolar point as a function of heliocentric distance, we are approximating the maximum differential stresses within the nucleus (see *Figure 3*).

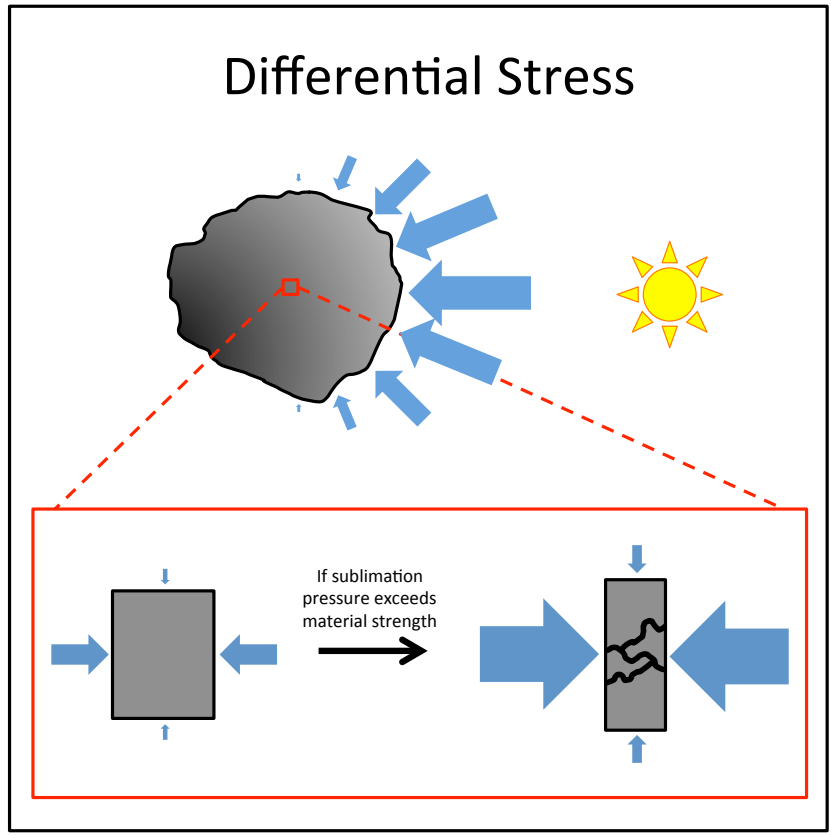

**Figure 3**: *Schematic of how Sublimation Pressure Induces Differential Stresses.* (top) dynamic sublimation pressure acts upon the sunward hemisphere of the nucleus. Sublimation pressure peaks at the subsolar point, but drops off to zero toward the limb. As the nucleus approaches the Sun, the sublimation pressure increases. (bottom inset) We illustrate the stresses acting on a parcel of material within the nucleus after subtracting off the hydrostatic pressure. The distribution of the sublimation pressure acting on the surface of the nucleus induces unequal stresses on the parcel of material, with stresses greatest along the comet-Sun axis. As the nucleus approaches the Sun, the stresses on the parcel grow. If the difference in stresses between the maximum stress and minimum stress axis (the differential stress) exceeds the strength of the material, then the parcel fails and fragments.

Gundlach et al. (2012) proposed a related mechanism, in which a sublimation pressure that pressed equally on all parts of the nucleus may have allowed Comet C/2011 W3 (Lovejoy) to survive through its perihelion of 1.2 $R_\odot$. Within ~10 $R_\odot$ of the Sun, the coma of a comet with a ~1 km nucleus becomes optically thick (Drahus et al. 2014), causing light of equal intensity to fall upon all parts of the nucleus, which results in a uniform sublimation pressure being exerted on all parts of its surface. Unlike our proposed mechanism, such a phenomenon would generate no new differential stresses within the interior of the nucleus. However, it would induce a confining pressure on its surface, which can increase the strength of porous, granular materials (Alkire & Andersland, 1973). If this increase in strength were sufficiently large, then volatile sublimation near the Sun could allow C/2011 W3 (Lovejoy) to resist the strong solar tidal forces that exist within the Roche Limit that would otherwise disrupt the nucleus (Gundlach et al. 2012).

The Whipple model for ice sublimation (Whipple, 1950), combined with our model of ISON as a sublimating sphere of ice 680 m in radius (Lamy et al. 2014), predicts a mass loss rate from Comet ISON's nucleus for $H_2O$ at 214 $R_\odot$ (1 AU) of $q_{water}$=2.75x10$^{28}$ s$^{-1}$, in agreement with the observed production rate of $q_{water}$ = 2.30(±0.71)x10$^{28}$ s$^{-1}$ (Combi et al. 2014). Measurements of Comet ISON's *Afρ* parameter as a function of aperture radius (*ρ*) flattened out and approached a constant value as ISON approached the Sun, suggesting that icy grains ceased to contribute significantly to ISON's volatile production by late October (Knight & Schleicher, 2015). We therefore find that such close agreement between the expected and measured production rates generally support our assumption that the entire illuminated hemisphere is sublimating.

Although Combi et al. (2014) deconvolved the observations with a model to obtain a daily average water production rate, their observed production rate of $q_{water}=1.99(\pm0.32) \times 10^{28}$ s$^{-1}$ at 0.98 AU is consistent with the measured rate of $q_{water} = 1.6 \times 10^{28}$ s$^{-1}$ ($\pm 25\%$) at 0.98 AU (Bodewits et al. 2013), and their observed production rate of $q_{water}=1.79(\pm0.35) \times 10^{28}$ s$^{-1}$ at 0.88 AU is within a factor of 2 of $q_{OH} = 8.14(\pm2.31) \times 10^{27}$ s$^{-1}$ at 0.89 AU (Opitom et al. 2013a). These observations, which demonstrate remarkable agreement across various instruments, are consistent with a highly active, intact nucleus.

However, after November 12$^{th}$, the amount of active surface required to match the observed H$_2$O production increased permanently by a factor of ~25, implying that the nucleus had then disrupted into a swarm of fragments (Combi et al. 2014). This is consistent with the observation of arc-like wings in the coma of ISON, which suggest the presence of multiple fragments (Boehnhardt et al. 2013). Other analysis determined that the radius of ISON's nucleus (or nucleus fragments) decreased too much during this event to be solely the result of sublimative surface erosion, further implying a disruption event at 145 $R_\odot$ (Steckloff et al. 2015). We therefore interpret this first event to be the complete breakup of the nucleus into a swarm dominated by large fragments ~100 m in radius (see Discussion section). The swarm (or specific large fragments within it) was later observed to undergo two further significant disruption events on November 21$^{st}$ and November 26$^{th}$ Knight & Battams, 2014; Steckloff et al. 2015).

Escaping fragments and lofted grains do not directly contribute to the reaction force on the comet's nucleus because their velocity is so slow near the nucleus relative to the sublimating gases that they carry a negligible amount of momentum away from the nucleus. However, they can reflect some fraction of the sublimated gas molecules back onto the nucleus, further increasing the dynamic pressure. This effect can only increase the peak dynamic pressure by a factor of $\pi$ (in the unlikely limit that every gas molecule bounces indefinitely between the nucleus and icy grains), to equal the gas vapor pressure. We adopt the conservative stance of neglecting this uncertain (but positive) backpressure, which can only add to the dynamic sublimation pressure, and which will introduce only small errors into our estimate.

## 3 Results:

Motivated by observations of high $H_2O$ production (Combi et al. 2014; Opitom et al. 2013a, 2013b, 2013c), we assume that volatile sublimation is dominated by $H_2O$ as ISON approached perihelion. We compute the maximum dynamic $H_2O$ sublimation pressure (and thus estimate the bulk cometary unconfined crushing strength) when Comet ISON disrupted at heliocentric distances of 36, 88, and 145 $R_\odot$ (Combi et al. 2014; Boehnhardt et al. 2013; Knight & Battams, 2014; Steckloff et al. 2015)]. We find strengths of 9, 1, and 0.5 Pa, respectively, for the case where ISON has a bond albedo of 0. If we instead assume a bond albedo of 0.5, we find strengths of 4, 0.6, and 0.2 Pa respectively (see *Figure 4*). These strengths are comparable to estimates of the strengths of Jupiter Family Comets (JFCs) (Asphaug & Benz, 1996; Bowling et al. 2014; Melosh, 2011; Sekanina & Yeomans, 1985; Thomas et al. 2015). If Comet ISON's true bond albedo is between these two values, then the maximum dynamic pressure and bulk unconfined crushing strength estimates will also lie between the corresponding values. Such a hierarchy of strengths is consistent with studies of the strength of geologic materials, which depend inversely on the size of the sample (Brace, 1961), and is consistent with evidence suggesting that comet nuclei are composed of pieces that are heterogeneous in strength (Sekanina, 2003). The lowest of these strength estimates (0.2 and 0.5 Pa depending on bond albedo) corresponds to the first disruption event (at 145 $R_\odot$), and therefore represents the bulk unconfined crushing strength of ISON's intact nucleus (prior to any significant fragmentation). The higher strength estimates correspond to the later disruption events at 88 $R_\odot$ and 36 $R_\odot$, and therefore represent the strengths of fragments of ISON's nucleus.

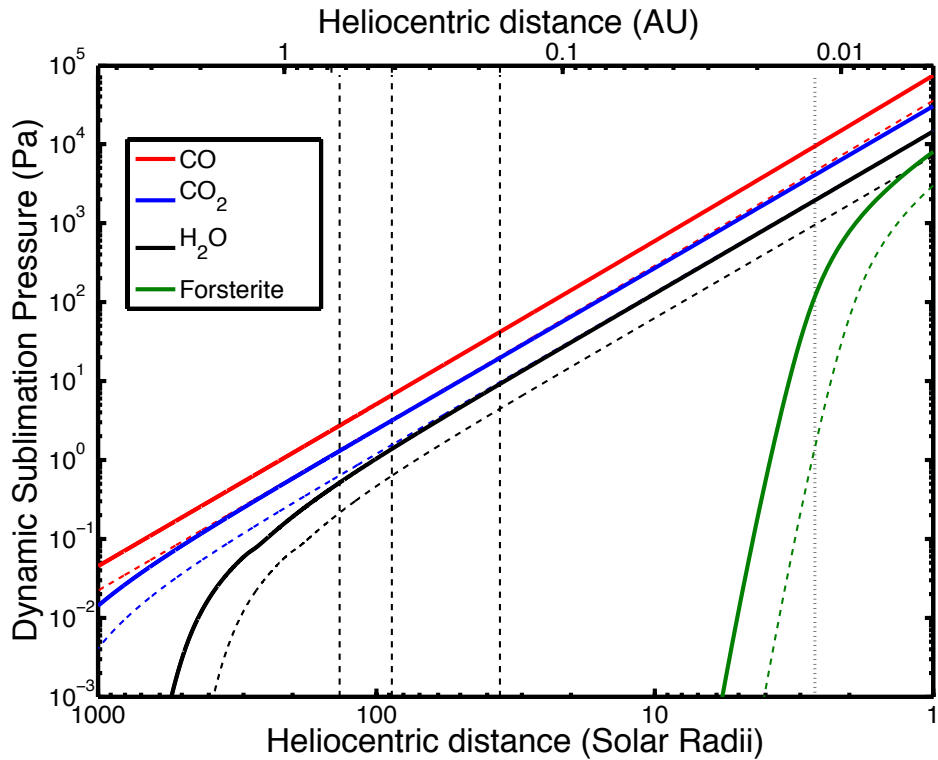

**Figure 4**: *Dynamic Gas Sublimation Pressures for Major Volatile Species.* A plot of dynamic gas pressures for pure $H_2O$, $CO_2$, and CO as a function of heliocentric distance, measured in both Solar Radii ($R_\odot$) and Astronomical Units (AU). We include the mineral fosterite (Nagahara et al. 1994) as a proxy for refractory cometary materials, which only becomes dominant in the absence of volatiles very near the Sun. Solid curves denote sublimation pressures if the nucleus has zero bond albedo while the dashed curves

are for an assumed bond albedo of 0.5. For a bond albedo between these two values, the sublimation pressure will lie between these two curves. The thin, dashed vertical lines at 36, 88[2], and 145[1] $R_\odot$ mark where Comet ISON disrupted into fragments (Combi et al. 2014; Knight & Battams, 2014), while the dotted line at 2.66 $R_\odot$ denote Comet ISON's perihelion distance.

## 4  Discussion:

After a fragmentation event, the size of the resulting fragments may have an observable effect on the motion of the comet or morphology of the nucleus. The sublimation pressure acting on the illuminated surfaces of the nucleus provides a net antisunward force, with the net motion of the nucleus dependent on this sublimation force and the solar gravitational force. Since the sublimation force depends on surface area, while the gravitation force depends on volume, larger bodies (smaller surface-area-to-volume ratio) are less susceptible to the sublimation force than smaller bodies (larger surface-area-to-volume ratio). Therefore, if the nucleus produced fragments of substantially unequal sizes, smaller fragments would appear to drift antisunward of the larger fragments, which would cause the central condensate of the comet's coma to elongate and even break up. However, Comet ISON maintained a strong central condensate (a compact region of peak coma brightness) up until only a few hours before perihelion (Knight & Battams, 2014; Opitom et al. 2013b, 2013c), and this central condensate only began to noticeably elongate a few days before perihelion (Steckloff et al. 2015). Thus, either the first fragmentation event broke Comet ISON into a swarm of equally sized fragments, or into differently sized fragments that were still each large enough to limit the relative drift between fragments and the resulting observable changes to the morphology of the coma.

Steckloff et al. (2015) conducted a preliminary study to estimate the sizes of the dominant fragments of Comet ISON. They measured the deviation of Comet ISON's position using the SCUBA-2 instrument on the James Clerk Maxwell Telescope from JPL Horizon's ephemeris solution #53, and estimated fragment sizes by assuming that this deviation is entirely due to $H_2O$ sublimation pressure. From this, they determined that the first fragmentation event reduced the effective radius of Comet ISON from an approximately 680 m for the intact nucleus to fragments on the order of ~100 m. Such fragments would require approximately half of a week to traverse a single pixel of the SCUBA-2 instrument and a few days more for the larger pixels of the TRAPPIST telescope. This provides a rough estimate of the timescale over which coma morphology would noticeably elongate from the release of a single fragment from a much larger parent nucleus. This timescale would be longer if the fragments are closer in size, since they would drift together. Since no change in coma morphology was detected during the 9 days between the first and second fragmentation events, it is unlikely that ISON only released a single ~100 m fragment from the nucleus during the first fragmentation event. Rather, it is more likely that the first fragmentation event broke up ISON's nucleus into a swarm of large fragments with radii on the order of ~100 meters.

Because the coma may have started to elongate between the second and third fragmentation events, it is unclear whether the second fragmentation event was the result of a single fragment or multiple fragments disrupting. However, the elongation of the central condensate after the third fragmentation event (Steckloff et al. *in prep.*) suggests that a large range of fragment sizes were present after the third fragmentation event.

*4.1 Supervolatiles and Amorphous Ice*

Samarasinha (2001) proposed that the buildup of pore pressure within the nucleus from the sublimation of super-volatile species could lead to its disruption. This mechanism requires that the thermal skin depth of the comet be large enough to reach pockets of deeply seated volatiles. The thermal skin depth ($h_{skin}$) describes the characteristic length scale over which the amplitude of a heat pulse conducting (without sublimating volatiles) into an infinite half-space of material with a fixed boundary location and temperature drops by a factor of *e*, and is given by the equation

$$h_{skin} = \sqrt{\mathcal{H}\tau} \qquad (13)$$

where $\mathcal{H}$ is a material's thermal diffusivity (typically on the order of $10^{-6}$ $m^2$ $s^{-1}$ for dense rocks or ice) and $\tau$ is the duration since the onset of the thermal pulse. The longer a material is exposed to a heat pulse, the deeper the heat can penetrate. The rate at which the thermal skin depth advances into a material is obtained by differentiating equation (13) with respect to time ($\tau$)

$$v_{skin} = \frac{dh_{skin}}{d\tau} = \frac{1}{2}\sqrt{\frac{\mathcal{H}}{\tau}} \qquad (14)$$

$$= \frac{\mathcal{H}}{2h_{skin}} \qquad (15)$$

Thus, as the time of exposure ($\tau$) and thermal skin depth ($h_{skin}$) increases, the rate of growth of the thermal skin depth ($v_{skin}$) decreases.

If the fixed-temperature boundary is receding at a constant rate, the thermal skin depth ($h_{skin}$) will either grow or shrink until $v_{skin}$ is equal to this rate of recession, and the thermal skin depth will maintain a fixed depth relative to the surface. However, because heat takes time to conduct from the surface to the thermal skin depth, the distance between the thermal skin depth ($h_{skin}$) and the receding surface will be less than what equation (13) provides. Also, the rate of surface recession on a comet nucleus is not constant, but rather accelerates as the nucleus approaches the Sun, which further reduces the distance between the surface and $h_{skin}$. Additionally, moving boundaries, changing boundary conditions, and sublimation make the actual temperature profile of a comet nucleus significantly more complicated than that which results from simple heat conduction. However, if we assume that the $H_2O$ sublimation front, whose temperature is largely determined by heliocentric distance, is some distance $h_{sub}$ below the surface of the nucleus and that $h_{skin}$ is measured from the sublimation front, then the quantity $h_{skin} + h_{sub}$ (computed using equations (13) and (15)) will be a conservative overestimate of Comet ISON's orbital thermal skin depth.

Because Comet ISON's activity occurred predominantly on the sunward hemisphere (Li et al. 2013c), the volatiles driving this activity had to respond to the day-night (diurnal) cycle of the nucleus, and could therefore be no deeper below the surface than a depth comparable to the diurnal skin depth. Based on a ~10.4 hour rotation period for the nucleus of Comet ISON (Lamy et al. 2014), the sublimation front of $H_2O$ ($h_{sub}$) is no more than ~20 cm below the surface. Since sublimation is a comet's dominant cooling mechanism in the inner Solar System, we estimate the rate of the sublimation front's recession into the nucleus at the time of the Lamy et al. (2014) observations by dividing the mass-loss rate equation (equation 2) by the bulk density of a typical comet, and find that it is on the order of ~$10^{-6}$ m/s. Noting that the rate of sublimation front recession and thermal skin depth recession ($v_{skin}$) are in equilibrium, we set $v_{skin}$ to ~$10^{-6}$ m/s, and find that $h_{skin}$ is on the order of ~0.5 m. Thus, $h_{skin} + h_{sub}$ is on the order of meters, and therefore cold, Oort Cloud conditions persist in the primordial materials of Comet ISON only a few meters at most below the surface of the nucleus.

Because the orbital thermal skin depth is so shallow, if the thermal wave were to reach a pocket of supervolatile ices or trigger the crystallization of amorphous ice, they would release fragments from the surface with sizes comparable to the orbital thermal skin depth. Thus, if the first fragmentation event were the result of the rapid sublimation of supervolatile species, one would expect to see an outburst that released debris up to an order of ~1 m in size, leaving the nucleus largely intact. If the nucleus is composed of amorphous water ice whose crystallization was triggered by the propagation of the thermal wave into the interior, the crystallization front will propagate into the amorphous ice until the cold interior of the nucleus absorbs the exothermic heat of the phase transition and quenches the crystallization process. Because the thermal wave is near the surface, the temperature gradient near the sublimation front is very steep (dropping to primordial temperatures over a distance on the order of the orbital thermal skin depth), and would quench the crystallization of the amorphous ice very quickly. Therefore, even if the exothermic crystallization of amorphous ice caused the first fragmentation event, one would still only expect to see an outburst that released similarly small debris.

Such small debris from a surface layer is inconsistent with the drastic reduction in the size of the nucleus after the first fragmentation event (Steckloff et al. 2015) and the observation of coma wings (Boehnhardt et al. 2013), which may indicate the presence of multiple large fragments. Additionally, such small debris would dissipate quickly, which is inconsistent with the sustained increase in water production (Combi, 2014). Therefore, while a direct application of the Samarasinha (2001) model may explain the disruption of highly thermally evolved comet nuclei, it appears that its direct application is inconsistent with the disruption of Comet ISON.

We cannot rule out a modification of the Samarasinha (2001) model, in which sublimating gases can penetrate into the pores of the nucleus and recondense (thus transporting heat into the cometary interior by releasing their heats of sublimation). If voids are present within the interior of the nucleus, then a sublimation front and thermal skin depth would be created within the walls of these voids akin to the situation at the surface. The Second Law of Thermodynamics limits the maximum temperature of the void walls achievable through this mechanism to the surface temperature of the nucleus (although the actual temperature would likely be much lower). Gas must be able to readily diffuse through the nucleus for a significant amount of heat to be transported into the cometary interior in this manner, which greatly restricts the ability of sublimating volatiles to build up a gas pressure as though the comet were a sealed vessel. As the walls of the void recede through sublimation, the thermal wave may encounter supervolatile ices or amorphous ice. The sublimation of supervolatiles within a void would produce pressures that could be no greater than those that would be present at the surface, but probably significantly less. If these low pressures lead to the destruction of the nucleus, then our strength estimates would be an upper bound to the strength of the nucleus. However, were the thermal wave to trigger the crystallization of amorphous ice, this exothermic phase transition could cause a very rapid buildup of gas pressure within the void, potentially faster than the gases may diffuse out, and could potentially lead to a catastrophic explosion of the nucleus. We therefore cannot rule out this modified mechanism. This mechanism requires special diffusive, compositional, and structural conditions to disrupt the nucleus, which seems less likely to lead to ISON's disruption than sublimation pressure at the surface. However, a detailed exploration of the relevant physics of diffusion, sublimation, and phase transitions is beyond the scope of this paper.

*4.2 Hydrostatic Pressure and Fragmentation Timescale*

Our crushing strength computation ignores the internal hydrostatic pressure due to self-gravity of comet ISON, which is up to ~10 Pa for a 680m spherical nucleus (Lamy et al. 2014) with a density of 400 kg m$^{-3}$ (Richardson & Melosh, 2013). If the nucleus were to uniformly disrupt in a single event, the dynamic sublimation pressure would have to overcome this overburden pressure in the comet's interior. In reality, the nucleus probably disrupted piecewise, in a process where the dynamic sublimation pressure first overcomes the crushing strength and disperses the material near the surface of the nucleus, where the hydrostatic pressure is low. This reduces the hydrostatic pressure throughout the remaining nucleus, where this process repeats until the entire cometary nucleus is dispersed. We estimate the timescale of this dispersion by computing the time needed for the surface of the comet to accelerate across the diameter of the nucleus from sublimation pressure alone, assuming typical cometary densities of around 400 kg m$^{-3}$ (Richardson & Melosh, 2013; Richardson & Bowling, 2014; Thomas et al. 2015). This results in a dispersion timescale for Comet ISON of only a few hours at 145 $R_\odot$, allowing us to ignore the effects of hydrostatic pressure and treat the cometary disruption effectively as an instantaneous event in the comet's orbit.

Our sublimation pressure disruption mechanism assumes that the nucleus is rotating slowly enough that the maximum dynamic sublimation pressure at the sub-solar region has enough time to fragment the nucleus before rotating significantly away from the sub-solar point and reducing the sublimation pressure on that area element. The critical timescale for fragmenting the nucleus is the amount of time needed for a crack, once started, to propagate across the nucleus. The growing tip of a crack travels at the Rayleigh surface wave velocity, which are typically on the order of ~100 m/s for granular materials, and higher for more coherent materials (Lawn & Wilshaw, 1975). Thus, the time needed for a crack to travel across the nucleus (and therefore the timescale of fragmentation) is on the order of a few seconds. Since the rotation period of a comet nucleus is limited to be no shorter than a few hours before fragmenting rotationally (Snodgrass et al. 2006; Pravec et al. 2006), the timescale of fragmentation is negligible and our assumption holds.

*4.3 Strengths of Other Comets*

We compare our crushing strength estimate to observationally constrained estimates of the bulk, tensile, and shear strengths of other comets, which are related to the bulk crushing strength by small factors on the order of unity (Price, 1968). The crushing strength of Comet ISON is consistent with Comet Shoemaker-Levy 9's bulk tensile strength of <6.5 Pa (Asphaug & Benz, 1996); Comet Brooks 2's bulk tensile strength of <2 Pa (Sekanina & Yeomans, 1985); within an order of magnitude of Comet Wild 2's shear strength of >17 Pa (Melosh, 2011); and Comet Churyumov-Gerasimenko's cohesive strength of ~2-16 Pa (Bowling et al. 2014), and tensile strength of <20 Pa (Thomas et al. 2015). Thus, if Comet ISON is representative of thermally unprocessed comets, then the low bulk strength of comets is a primordial property that is unaltered by thermal processing.

We consider other strength estimates of comets, and note that they are not applicable to our mechanism. The 1-10 kPa effective target strength of Comet 9P/Tempel 1 from the Deep Impact experiment (Richardson & Melosh, 2013) is a measurement of dynamic strength (which does not adhere to the weakest link model of material failure). Therefore, we expect this estimate to be several orders of magnitude larger than a measurement of static strength, which is applicable to our disruption mechanism. Comet Hyakutake's tensile strength was estimated to be ~100 Pa from the strength required to hold the comet together from rotational fragmentation (Lisse et al. 1999). However, this estimate assumed a bulk density for Comet Hyakutake of 100 kg m$^{-3}$, which is now known to be unreasonably low: a more typical cometary density of 270 kg m$^{-3}$ or greater allows the nucleus to be held together by gravity alone. Indeed the known rotation rates of JFCs and Kuiper Belt Objects are consistent with effectively strengthless bodies with densities less than 600 kg m$^{-3}$ (Snodgrass et al. 2006) in a manner analogous to the asteroid rubble pile "spin barrier" (Pravec et al. 2006).

All of these upper bounds of comet strength require that the nucleus structurally fail in some way. Thus, these strength estimates may be biased toward weaker nuclei, which would structurally fail more easily. Indeed, many comets survive perihelion passage despite having orbits that take them to smaller heliocentric distances than those corresponding to Comet ISON's fragmentation events (Bortle, 1991), consistent with stronger nuclei. If comets are effectively rubble piles held together by van der Waal's forces, then they may possess strengths similar to rubble pile asteroids of ~25 Pa (Sánchez & Scheeres, 2014). Such strengths would allow comet nuclei to survive the differential stresses induced by $H_2O$ sublimation to within 20 $R_\odot$ (0.1 AU) of the Sun. Thus, the survival/non-survival of near-Sun comets is consistent with different comet nuclei having strengths that span more than an order of magnitude.

Additionally, short-period comets with small perihelia (when compared to where ISON fragmented) may survive multiple orbits as a result of their unique dynamical and thermophysical evolution. Jupiter Family Comets like 2P/Encke and 96P/Machholz originate in the Kuiper Belt and Scattered Disk until an encounter with Neptune sends them into the Outer Planet region of the Solar System, where they are reclassified as Centaurs (Duncan et al. 2004). Typically, an encounter with Jupiter after a few million years (the dynamical lifetime of a Centaur) either ejects the object from the Solar System or sends it into the Jupiter Family of comets (Duncan et al. 2004). During this inward migration process, a Jupiter Family Comet is also undergoing thermophysical evolution. As its orbit evolves ever closer the Sun, the comet loses volatile ices through sublimation, which may result in the build up of a lag deposit (or dust mantle) on its surface. These deposits are very good insulators (Gulkis et al. 2015; Davidsson et al. 2013; Groussin et al. 2013; Lisse et al. 2005; Lamy et al. 2008), and even a thin coating would restrict volatile sublimation to a small fraction of the surface. Therefore, when this inhibited sublimation activity is averaged over the surface, we expect JFCs to experience significantly lower sublimation pressures than the pristine icy surfaces that we have modeled in this work. Thus, the survival of JFCs with small perihelia is consistent with our work, even without allowing for larger material strengths.

## 5 Conclusions

We have shown that existing mechanisms of comet disruption have difficulty explaining Comet ISON's fragmentation. We proposed a new mechanism of comet disruption in which sublimating gases exert a dynamic pressure on the sunward hemisphere of a nucleus and induce differential stresses within the nucleus, which may fracture and fragment the nucleus if they exceed its material strength. Using a versatile thermodynamic model of volatile sublimation, we find Comet ISON has a material strength similar to JFCs. For the case that the nucleus of Comet ISON has a bond albedo of 0, we estimate its bulk unconfined crushing strength to be 0.5 Pa, and the bulk unconfined crushing strength of resulting fragments at 1-9 Pa. If Comet ISON's nucleus has a bond albedo of 0.5, then these strength estimates drop to 0.2 Pa for the intact nucleus and 0.6-4 Pa for its fragments.

**Competing Financial Interests**

The authors of this Letter have no competing financial interests regarding this work.

Correspondence should be addressed to Jordan K. Steckloff

**Acknowledgements**

This research project was internally funded though the startup funds of H. Jay Melosh (Purdue University). We thank Nalin Samarasinha and an anonymous referee for helpful comments.